\documentclass[letterpaper]{JHEP3} 
\usepackage{amsmath}
\usepackage{epsfig}
\usepackage{cite}
\usepackage{pstricks,pst-node}
\makeatletter
\let\default@color=\current@color
\makeatother
\newcommand{\beq}{\begin{equation}}
\newcommand{\eeq}{\end{equation}}
\newcommand{\bea}{\begin{eqnarray}}

\newcommand{\eea}{\end{eqnarray}}
\newcommand{\beas}{\begin{eqnarray*}}
\newcommand{\eeas}{\end{eqnarray*}}

\newcommand{\infint}{{\int_{-\infty}^\infty \hspace{-.3cm}dy\,}}
\newcommand{\tinfint}{{\int_{-\infty}^\infty \hspace{-.3cm}d\tilde{y}\,}}
\newcommand{\sint}{\sum\hspace{-0.35cm}\int\hspace{0.2cm}}
\newcommand{\sintf}{\sum_{n\neq0}\hspace{-0.45cm}\int\hspace{0.1cm}}
\newcommand{\sintfm}{\sum_{m\neq0}\hspace{-0.55cm}\int\hspace{0.1cm}}
\newcommand{\sintfmneqn}{\sum_{\genfrac{}{}{0pt}{1}{m\neq0}{m\neq n}}\hspace{-0.55cm}\int\hspace{0.1cm}}
\newcommand{\sinthmneqn}{\sum_{\genfrac{}{}{0pt}{1}{m\neq1}{m\neq n}}\hspace{-0.55cm}\int\hspace{0.1cm}}
\newcommand{\dsintf}{\sum_{n,m\neq0}\hspace{-0.63cm}\int\hspace{0.1cm}}
\newcommand{\sinth}{\sum_{n\neq1}\hspace{-0.45cm}\int\hspace{0.1cm}}
\newcommand{\sinthm}{\sum_{m\neq1}\hspace{-0.55cm}\int\hspace{0.1cm}}
\newcommand{\dsinth}{\sum_{n,m\neq1}\hspace{-0.63cm}\int\hspace{0.1cm}}

\title{Effective action of a five-dimensional domain wall}

\author{Y. Burnier\\ 
Faculty of Physics, University of Bielefeld, D-33501 Bielefeld, Germany\\
E-mail: \email{yburnier@physik.uni-bielefeld.de}}

\author{K. Zuleta\\
Department of Physics, University of Ioannina, GR-45110 Ioannina, Greece\\
E-mail: \email{kzuleta@uoi.gr}}

\abstract{We calculate the four-dimensional low-energy effective action for the perturbations of a two-scalar domain wall model in five dimensions. Comparison of the effective action to the Nambu-Goto action reveals the presence of an additional coupling between the light scalar field and the massless translation mode (branon excitation), which can be written in terms of the curvature scalar of the induced metric. We comment on the impact of this interaction to branon physics.} 


\preprint{}

\begin{document}

\section{Introduction\label{introduction}} 

Since a few years, the  braneworld idea put forward already in the eighties~\cite{misha_1,akama} has made a spectacular comeback becoming a vital ingredient of higher-dimensional theories. In essence, the idea is that by confining the Standard Model fields on a brane, a hypersurface embedded in a higher-dimensional space-time, one can obtain a low energy theory which looks four-dimensional without the need to compactify the extra dimensions. 

In the original scenarios, branes are modelled as defects of a higher-dimensional field theory, domain walls~\cite{misha_1} or  strings~\cite{akama}. Light modes of the higher-dimensional fields, localized on the defect, are interpreted as four-dimensional particles. Using these scenarios, one can naturally build theories with scalars and fermions living on the defect - e.g. in~\cite{misha_1} fermions were incorporated through Yukawa  interaction with the scalar creating the domain wall. Localization of gauge fields is, however, rather problematic (especially for the non-abelian fields, because of the difficulty to achieve charge universality), although some ideas towards the resolution of this problem have been pursued in~\cite{dvali_shifman}.

The recent revival  of the braneworld idea was stimulated by new achievements within the string theory, in particular the discovery~\cite{d-branes} of Dirichlet branes (or D-branes), solitons of the string theory on which open strings can end and which have the property of localizing gauge fields (as well as fermions and bosons interacting with them). It has also been realized that gravity can be incorporated into the braneworld picture, along one of two alternative approaches: The first idea, put forward in the ADD models~\cite{add_1,add_2}, was simply to use compactification to make gravity four-dimensional (at distances exceeding the compactification scale), but under the assumption that only gravity can probe the extra dimensions, all the Standard Model fields being confined on a brane by some mechanism or other. The advantage, compared to the classic Kaluza-Klein picture, was to allow for much bigger size of the extra dimensions, as it was only constrained by the small-distance Newton's law experiments. In this scenario the gravitational backreaction of the brane on the bulk geometry is supposed to be negligible. The second approach, put forward in the RS models~\cite{rs1,rs2}, is to trap gravity itself on a brane through  warping of space-time. Indeed, it was shown in~\cite{rs1, rs2} that when a brane is embedded in a five-dimensional anti-de Sitter space-time, the spectrum of excitations of the five-dimensional graviton contains a localized zero mode, which can be seen as a four-dimensional graviton. 

Infinitely thin branes appearing in the RS or ADD models can be thought as approximations of smooth structures, thick branes, defects of some higher-dimensional field theory. Several regularized versions of the non-compact RS2 model have been constructed~\cite{kehagias,gremm_1,gremm_2, massimo_1}, along very much the same lines as in~\cite{misha_1}, using smooth gravitating domain walls to generate warped geometry. As for the ADD models, their physics has been studied using an effective description of the brane and its four-dimensional field theory (without actually dealing with the question of how the brane appears). This approach, motivated by the lack of a truly realistic model within which the Standard Model would be localized on a brane (or several branes) -- be it field-theoretical model or string theory model\footnote{Note that an impressive progress has been made in the recent years in constructing semi-realistic intersecting D-brane models, see e.g. the reviews~\cite{d_branes_rev, d_branes_rev_2} and the references therein.} --  has been developped in~\cite{sundrum_eff}. It relies on the four- and five-dimensional covariance principles to construct the action for a field theory on a thin brane, with Nambu-Goto~\cite{ng_dirac} action as a foundation. This approach proved to be a very powerful tool to study the physics of excitations of the ADD-type models, the spectrum of which contains besides the Kaluza-Klein gravitons, also brane's own excitations: exotic scalar particles called branons~\cite{kugo_ng,branon_dyn}. Branons can be interpreted as Goldstone bosons appearing as a result of breaking of the isometries of the extra space by the presence of the brane. It was shown~\cite{kugo_grav} that when the brane tension in much smaller than the fundamental scale of gravity, massive Kaluza-Klein modes decouple from the branons, allowing (at least in principle) a low-energy description of the dynamics of these particles. It has been argued~\cite{branon_dyn} that when the isometries being broken are not exact, branons get a mass. Massive branons has been advocated as dark matter candidates~\cite{dark_branons,dark_branons_2}, as they are supposed to be stable and interact very weakly with the Standard Model fields. It was suggested~\cite{branons_exp,branons_lhc} that they could be observed in collider experiments (see~\cite{branons_lep,branons_lep_2} for LEP limits). 

The aim of the present work is to determine the four-dimensional low energy effective field theory on a brane in an explicit five-dimensional model and to verify how close it is to the Nambu-Goto action. Our main goal is to investigate the form and the strength of branon interactions.  We do not aspire to present a realistic braneworld scenario and for the sake of simplicity, our study will be based on a very basic setup, where the brane is modelled as a domain wall in a five-dimensional Minkowski space-time and is populated only by a light scalar field and a massless branon. We will ignore the issue of producing four-dimensional gravity, which could either appear as induced gravity~\cite{dgp_1,dgp_2}  or  produced through compactification of extra dimension(s), and concentrate on the interactions of the scalar with the branon excitation. Given that both fields are localized on the brane and in view of the results of~\cite{kugo_grav}, we do not expect their interactions to be influenced by the size of the extra dimensions.   

The paper is organized as follows: in section~\ref{section_ng}, we briefly remind the Nambu-Goto description of a brane. In section~\ref{section_domain_wall}, we present the five-dimensional domain wall  model with two scalars and determine its spectrum of excitations. Section~\ref{section_effective_action} is devoted to the calculation of the four-dimensional low energy effective action for this model and its comparison with the Nambu-Goto action. In section~\ref{section_physical_implications} we discuss physical implications of the additional branon interaction appearing in our action.

\section{Nambu-Goto action for the brane \label{section_ng}}

The effective action for the field theory on a thin brane was constructed, by symmetry considerations, in~\cite{sundrum_eff}. Let us remind briefly this approach in order to set up the notations and to facilitate the analysis of our results.   

Let us consider a thin brane embedded in a five-dimensional bulk which we will take to be a Minkowski space-time -- a simplifying assumption, which is justified in the absence of any other fields in the bulk and when the  brane tension~$\tau$ is small compared to the fundamental gravity scale, $\tau\ll M_P^4$, so that we can neglect the backreaction of the brane on the bulk geometry.   
Let us denote the bulk coordinates as~$X^M$,~$M=0,\dots, 4$ and let the internal coordinates on the brane be~$x^\mu$~,~$\mu=0,\dots, 3$. The brane is then a hypersurface which can be described through parametric equations~$X^M=Y^M(x)$. If we choose the gauge~$Y^\mu(x)=x^\mu$, $\mu=0,\dots,3,$ the induced metric reads\footnote{We use the ``mostly minus'' convention for the signature of the metric, that is $\eta_{MN}=\mathrm{diag}(1,-1,-1,-1,-1)$.} :   
\begin{equation}
g_{\mu\nu}=\partial_\mu Y^M\partial_\nu Y^N \eta_{MN}=\eta_{\mu\nu}-\partial_\mu Y\partial_\nu Y\ ,
\end{equation}
where $Y(x)=Y^4(x)$ denotes the position of the fluctuating brane in the bulk. From the four-dimensional point of view,~$Y(x)$ is a dynamical massless scalar field, a so-called \emph{branon}:  Goldstone boson appearing as a consequence of the breaking of translation symmetry along the extra dimension by the presence of the brane.  

The prescription for constructing the effective action describing a four-dimensional field theory on the brane is that it should obey the five-dimensional covariance, as well as be invariant under four-dimensional coordinate transformations. As a result, the effective action is basically the usual action for a matter field in a curved space-time (the metric of which is the induced metric~$g_{\mu\nu}$). In particular, supposing that the only matter field on the brane is a scalar~$v_1$ with a potential $V(v_1)=\tilde{m}_1^2v_1^2/2+\lambda_4 v_1^4/4$ (which is the case we will be concerned with in what follows), the brane action reads:  
\begin{equation}
S^{\text{scalar}}_{\text{NG}}= \int d^4 x \sqrt{g}{\left\{ -\tau + {\cal{L}}_{\text{scalar}}\right\}=\int d^4 x \sqrt{g}\left\{-\tau+\frac12 g^{\mu\nu}\partial_\mu v_1\partial_\mu v_1 -V(v_1)\right\}} .  
\label{NG_1}  
\end{equation}
The leading constant term in the action is the Nambu-Goto term,~$\tau$ being the brane tension. 
As it is transparent from~(\ref{NG_1}), branons interact with the matter fields localized on the brane only through the induced metric - their interactions are therefore only derivative\footnote{When gauge fields are present, branons can also have interactions of other type, see~\cite{kugo_ng}.}. These interactions can be seen more explicitly in the low energy expansion:
 \begin{eqnarray}
S^{\text{scalar}}_{\text{NG}}&=&\int d^4 x {\left\{-\tau+\frac12 \partial^\mu\tilde{Y}\partial_\mu\tilde{Y}+\frac{1}{8\tau} {\left(\partial^\mu\tilde{Y}\partial_\mu\tilde{Y}\right)}^2+\frac12\partial^\mu v_1\partial_\mu v_1 -\frac12 \tilde{m}_1^2 v_1^2 -\frac{\lambda_4}{4}v_1^4 \right.} \nonumber \\
 && \hphantom{\int d^4 x \left\{\right.}{\left.+\frac{1}{2\tau}\partial^\mu\tilde{Y}\partial^\nu\tilde{Y}\partial_\mu v_1\partial_\nu v_1 -\frac{1}{4\tau} \partial^\mu\tilde{Y}\partial_\mu\tilde{Y}\left[\partial^\alpha v_1\partial_\alpha v_1-\tilde{m}_1^2v_1^2\right]+\dots\right\}} ,  
\label{NG_exp}
\end{eqnarray}
where we have rescaled $Y(x)$, introducing: 
$$
\tilde{Y}(x)\equiv\sqrt{\tau}Y(x)
$$ 
which carries the mass dimension of a four-dimensional scalar field, and used:
$$
g^{\mu\nu}=\eta^{\mu\nu}+\frac{1}{2\tau}\partial^\mu\tilde{Y}\partial^\nu\tilde{Y}+\dots
$$
and 
$$
\sqrt{g}=1-\frac{1}{2\tau}\eta^{\mu\nu}\partial_\mu\tilde{Y}\partial_\nu\tilde{Y}-\frac{1}{8\tau^2}\left(\partial^\mu\tilde{Y}\partial_\nu\tilde{Y}\right)^2+\dots
$$
   
Our goal will be to determine the effective action in a domain wall model containing two scalar fields and to verify how close this action is to the action~(\ref{NG_1}), to which we will from now on loosely refer as to the Nambu-Goto action.

Depending on the accuracy of the calculations, we might expect in our action corrections coming from the finite  width of the domain wall. Such corrections to the Nambu-Goto action were investigated in~\cite{ruth_garfinkle_1, ruth_garfinkle_2,ruth_carter} using Gauss-Coddazzi formalism and were found to be proportional to the brane's curvature (both intrinsic and extrinsic). More explicitly, it was found in~\cite{ruth_carter} that for the case of the $\lambda\Phi^4$ domain-wall model, the action of the brane reads (in the absence of matter on the brane):
\begin{equation}
\label{width_corrs}
S_{\text{brane}}=\int d^4 x\sqrt{g}\, {\left\{-\tau \left[1+\frac{\pi^2-6}{24 a^2}R -\frac{1}{3a^2}K^2\right]+{\cal{O}}\left(\frac{1}{a^4}\right)\right\}} ,
\end{equation}
where~$R$ denotes the curvature scalar and~$K$ the extrinsic curvature of the brane and~$a^{-1}$ is the width of the domain wall. 
  
\section{Setup: two-field domain wall\label{section_domain_wall}}
As an explicit field-theoretical (toy) model of a braneworld, we will use a  five-dimensional domain wall model with two scalar fields, described by the following action\footnote{This model has also been considered in~\cite{volkas} where the spectrum of its perturbations has been determined.}:
\begin{eqnarray}    
\label{Phi-Xi_action}
S=\int d^4x dy && \hspace{-0.2cm}\left[\frac{1}{2}\eta^{MN}\partial_M\Phi\partial_N\Phi+\frac{1}{2}\eta^{MN}\partial_M\Xi\partial_N\Xi  -V(\Phi,\Xi) \right] \ ,\nonumber \\
V(\Phi,\Xi)=&& \hspace{-0.1cm}\frac{\lambda}{4}\left(\Phi^2-v^2\right)^2 
+\frac{\tilde\lambda}{4}\Xi^4 +\frac12M^2\Xi^2 
+\frac12\alpha(\Phi^2-v^2)\Xi^2{\vphantom{\frac12}}  \ ,
\end{eqnarray}
 Provided~$\lambda\tilde\lambda v^4>(\alpha v^2-M^2)^2$, the system has a degenerate ground state $(\Phi_{GS}=\pm v, \Xi_{GS}=0)$ and we can therefore set up a domain wall interpolating between the two vacua\footnote{Condition~$\lambda\tilde\lambda v^4>(\alpha v^2-M^2)^2$ ensures that the potential is positive everywhere except at $(\Phi_{GS},\Xi_{GS})=(\pm v,0)$ where it vanishes. The detailed form of the potential depends on the choice of parameters and for instance for $M^2<\alpha v^2$ additional local minima (with higher energy) may appear.}. It can be easily verified that the kink configuration:
\begin{equation}
(\Phi_K,\Xi_K)=(v\tanh(ay),0) 
\label{kink}
\end{equation}
with $a^2=\lambda v^2/2$ is always a solution to the classical equations of motion. 

As per usual, excitations of~$\Phi$ and~$\Xi$ localized on the defect will play the role of the four-dimensional fields. In order to determine the field content of the four-dimensional theory we consider perturbations around the kink configuration, which can be written:
\begin{eqnarray}
\Phi(x,y)&=&\Phi_K(y)+\phi(x,y)=\Phi_K(y) +\smash{\sum_{n}\hspace{-0.45cm}\int\hspace{0.2cm}}f_n(y)u_n(x)\ ,\nonumber \\[2ex]
\Xi(x,y)&=&\xi(x,y)=\smash{\sum_{k}\hspace{-0.45cm}\int\hspace{0.2cm}} h_k(y)v_k(x)\ , \label{expansion_1}
\end{eqnarray}
where we use the sign~$\sint$as a shorthand indicating both summing over the discrete states of the spectrum and integration over the continuum. The modes $u_n(x)$ and $v_n(x)$ satisfy the  four-dimensional Klein-Gordon equations (and therefore can be interpreted as four-dimensional scalar fields) and~$f_n(y)$ and~$h_n(y)$ are the wavefunctions determining the localization of the modes on the brane ($f_n(y)$ being the familiar normal modes of the kink). These satisfy the following Schr\"odinger-like equations:   
\begin{equation}
\label{normal_modes}
\begin{cases}
\displaystyle -\partial_y^2f_n+ \left(4a^2-\frac{6a^2}{\cosh^2(ay)}\right)f_n=m_n^2 f_n \\
\displaystyle -\partial_y^2 h_k  +\left(M^2-\frac{\alpha v^2}{\cosh^2(ay)}\right)h_k =\tilde m_k^2 h_k\ ,
\end{cases}
\end{equation}
where the eigenvalues~$m_n^2$ and~$\tilde{m}_n^2$ are masses of the four-dimensional fields (squared).

As expected, the lightest state in our model is the zero mode~$u_0(x)$ with  wavefunction:
\begin{equation}
\label{zero_0}
f_0(y)=\frac{\sqrt{3a}}{2\cosh^2(ay)} \ .
\end{equation}
It is the {\emph{translation mode}} of the kink, appearing as a result of spontaneous breaking of the translation invariance along the extra dimension by the presence of the brane. It should be noted that~(\ref{kink}) is one among the infinity of kink configurations of equal energy. Indeed, as the Lagrangian in~(\ref{Phi-Xi_action}) is translationaly invariant, the equations of motion are satisfied by 
$$(\Phi_K^{(y_0)},\Xi_K^{(y_0)})=(v\tanh(a(y-y_0)),0)$$
 for an arbitrary $y_0$. Naturally, being $y$-dependent any such kink configuration breaks the translational symmetry. Under an infinitesimal translation,~$\Phi_K$ changes by:
$$
\Delta\Phi_K=\Phi_K(y-\delta y_0)-\Phi_K(y)\approx-\delta y_0 \Phi'_K(y)\ ,
$$ 
where $\Phi'_K(y)$ is nothing else but the eigenfunction~$f_0$ of the zero mode (up to a normalization constant). The fact that $u_0$ is massless reflects the fact that a displacement leaves the kink's energy unchanged.  

The rest of the spectrum of perturbations of $\Phi$ consists of a heavy mode of mass $m_1^2=3a^2$ and of the continuum of modes which are not localized on the brane. 

As for the spectrum of~$\Xi$, the number of modes localized on the brane and their masses depend on the parameters of our model. The lowest localized mode of~$\Xi$ is~$v_1(x)$ with wavefunction:
\begin{equation}
\label{massive_0}
h_1(y)=\frac{N_1\sqrt{a}}{\cosh^{\sigma}(ay)}
\end{equation}
where
\begin{equation}
\label{sigma}
\sigma=-\frac12+\frac12\sqrt{1+8\frac{\alpha}{\lambda}} \quad \mbox{and} \quad N_1=\left(\int_{-\infty}^\infty \frac{dy\, a}{\cosh^{2\sigma}(ay)}\right)^{-1/2}=\sqrt{\frac{\Gamma\left(\sigma+\frac12\right)}{\sqrt{\pi}\Gamma(\sigma)}}\ .
\end{equation}
Mass of $v_1(x)$ is given by
$$
\tilde m_1^2=-\sigma^2 a^2+M^2 
$$
and therefore it can be made small (compared to~$a$, the heavy scale of the model) by setting
$$
M^2=\sigma^2 a^2 +\frac{\lambda v^2}{4}\epsilon^2\ , 
$$
with $|\epsilon|\ll 1$. We have then 
$$
\tilde{m}_1^2=\frac{\lambda v^2}{4}\epsilon^2\ .
$$ 
Whether or not there are other localized modes in the spectrum of $\Xi$ depends on the value of~$\sigma$, but if they are present, their masses are necessarily large compared to $\tilde m_1$. We thus have a gap between the light and the heavy modes which can be made arbitrarily large by tuning~$\epsilon$.

To resume, the particle content of our low energy four-dimensional theory will consist of two scalar fields -- one massless $u_0$ which is an intrinsic brane perturbation, and one massive~$v_1$, which will play the role of the ``matter field'' localized on the brane. The gap between the light modes and the heavy modes is proportional to~$a$, and therefore the larger~$a$ is, the wider the gap becomes. This feature will allow us to derive the four-dimensional low energy effective  action for the light fields as a series in powers of~$1/a$ (or, equivalently, in powers of~$\epsilon$). Additionally, as our calculations will be mainly classical, we need to work in the weak coupling regime in order to be able to neglect consistently quantum corrections\footnote{The first quantum corrections typically take the form of the one loop diagram with two $\lambda \Phi^4$ vertices. In five dimensions this diagram will be of order $\lambda ^2 a$. Imposing that this correction is small in comparison to just one $\Phi^4$ vertex gives $\lambda^2a \ll \lambda$.}. To this aim, we require that all the five-dimensional couplings are roughly of the same order and small (in dimensionless units), 
\beq
\lambda a\sim \tilde{\lambda} a\sim \alpha a \ll 1.\label{weakcoupling}
\eeq
The effects of the nonlinear couplings in effective action can therefore also be expanded in powers of these couplings. Note that in the following, we will use the fact that these couplings are of the same order and loosely refer to this as the expansion in powers of $\lambda a$. In order to ensure that the radiative corrections do not spoil the mass hierarchy of the modes, we will also assume~$\epsilon^2\gg\lambda a$.  
   
In fact, rather than using (\ref{expansion_1}) in order to derive the four-dimensional low energy effective action, it is judicious to replace~$u_0(x)$ with a \emph{collective coordinate} (see e.g.~\cite{rajaraman}) associated with the translation of the kink which we will denote~$Y(x)$\footnote{Essentially, introducing~$Y(x)$ amounts to promoting the constant~$y_0$ denoting the position of the center of the kink into a dynamical variable.}, and rewrite the perturbations around the classical kink configuration as follows:
\begin{eqnarray}
\begin{cases}
\displaystyle \Phi(x,y)=\Phi_K(y-Y(x)) + \sintf f_n(y-Y(x))\,u_n(x)\ ,\\
\displaystyle \Xi(x,y)= h_1(y-Y(x))\,v_1(x)+ \sinth h_k(y-Y(x))\,v_k(x) \ .\\[-.5ex]
\end{cases}
\label{perts}
\end{eqnarray}  
This parametrization has two advantages: first of all, the collective coordinate $Y(x)$ has a simple geometric interpretation as the transverse coordinate of the brane (hence our choice of notation) and therefore using the collective coordinate approach allows us to make a direct connection with the geometric framework in which the Nambu-Goto action is obtained (as described in section~\ref{section_ng}). The comparison between the  low energy effective action of the domain wall, which we will present in the next section and the Nambu-Goto action, eqns.~(\ref{NG_1}) and~(\ref{NG_exp}),  will be completely straightforward.  Secondly, expansion~(\ref{perts}) captures the Goldstone boson nature of the massless scalar present in our theory better than~(\ref{expansion_1}), yielding automatically only derivative interactions for the branon field~$Y(x)$. 

\section{Effective action\label{section_effective_action}}
Let us now proceed to determine the four-dimensional low energy effective action, which is constructed, as usual, as a series in the inverse of the heavy scale of the model (in our case,~$a$) and its small parameters, in our case~$\epsilon=\tilde{m}_1/a$ and~$\lambda a$ (let us remind that~\mbox{$\lambda a\ll 1$} is the weak coupling condition for our model). More specifically, our goal will be to determine the effective action at the tree level at the first order in coupling~$\lambda$. Our aim is to get more insight into the interactions of the zero mode (the branon field).  Comparison with the Nambu-Goto action~(\ref{NG_exp}) requires that we determine the action up to the order~${\cal{O}}(1/\tau)$.
\subsection{Formal expansion}
 In order to calculate the effective action, we first substitute the expansion~(\ref{perts}) into the action~(\ref{Phi-Xi_action}) and perform the integration over~$y$ (using the orthonormality of the wavefunctions~$f_n(y)$ and~$h_n(y)$). Unsurprisingly, this produces a four-dimensional action containing a plethora of terms, including interactions between the light fields, interactions of the light fields with one, two or three heavy fields and finally interaction terms involving only the heavy fields. More explicitly:
\begin{eqnarray}
S&=&\int d^4 x \left\{-\tau+\frac12 \partial^\mu\tilde{Y}\partial_\mu\tilde{Y}+\frac12\partial^\mu v_1\partial_\mu v_1 -\frac12 \tilde{m}_1^2 v_1^2
\smash{-\frac{{\lambda_4^{(0)}}\!\!\!}{4}\,v_1^4+\lambda_{(2,2)}^{(0)} v_1^2\,\partial^\mu\tilde{Y}\partial_\mu\tilde{Y} }\right. +\qquad\nonumber \\
&&\hphantom{\int d^4 x \left\{\right.}+\frac12\sintf \left[\partial^\mu u_n\partial_\mu u_n-m_n^2 u_n^2\right] 
+\left.\frac12\smash{\sinth \left[\partial^\mu v_n\partial_\mu v_n-\tilde{m}_n^2 v_n^2\right]} + \mathcal{L}_{\text{heavy}}^{\text{int}}\right\}
\label{action_exp}
\end{eqnarray}
 where 
 $\mathcal{L}_{\text{heavy}}^{\text{int}}$ contains the interactions between the heavy and the light modes and where again, as in section~\ref{section_ng}, we have replaced $Y(x)$ by the properly normalized field $\tilde{Y}(x)$:
$$\tilde{Y}(x)=\sqrt{\tau}\, Y(x)\ .$$
The leading term of our action: 
$$\tau=\int dy \, \Phi_K'^{\,2}=\frac43 a v^2=\frac{8a^3}{3\lambda}$$
is the tension of the brane (the energy density of the kink). 
 Let us notice that the weak coupling condition~(\ref{weakcoupling}) implies $\tau\gg a^4$. % 
The couplings of the interaction terms involving only the light fields can be expressed by the parameters of the five-dimensional model as follows:
\begin{eqnarray}
\lambda_4^{(0)}&=&{\tilde\lambda}\int dy\, h_1^4= {\tilde\lambda}a \frac{\Gamma^2(\sigma+\frac 1 2)\Gamma(2\sigma)}{\sqrt{\pi}\Gamma^2(\sigma)\Gamma(2\sigma+\frac 1 2 )}\label{lambda_4_0}\\
\lambda_{(2,2)}^{(0)}&=&\frac{1}{\tau}\int dy\, h_1'^2=\frac{\sigma^2}{(1+2\sigma)}\frac{a^2}{\tau}=\frac{3\sigma^2}{8(1+2\sigma)}\frac{\lambda}{a} \label{lambda_22_0}\ .
\end{eqnarray}
Finally, sorting the interactions between light and heavy modes by the number of heavy modes we have
\bea
 \mathcal{L}_{\text{heavy}}^{\text{int}}&=& \sintf J_n^{(1)} u_n+\frac12\dsintf{J}^{(2)}_{nm} u_n u_m+\dsintf{K}^\mu_{nm} u_n \partial_\mu u_m+ \nonumber \\
&&+\sinth\tilde{J}^{(1)}_n v_n+\frac12\dsinth\tilde{J}^{(2)}_{mn} v_n v_m +\dsinth\tilde{K}^\mu_{nm} v_n \partial_\mu v_m+ \nonumber \\ &&+ \sintf\sinthm \tilde{\tilde{J}}_{n m}^{(2)} u_nv_m+\dots 
 ,
\eea
where $J_n^{(1)}$, $\tilde{J}_n^{(1)}$, $J_{nm}^{(2)}$, $\tilde{J}_{nm}^{(2)}$, $\tilde{\tilde{J}}_{nm}^{(2)}$, $K^\mu_{nm}$ and~$\tilde{K}^\mu_{mn}$ denote some combinations of the light fields. Their exact form is at this stage inessential (see Appendix~\ref{app_heavy_couplings} for the explicit expressions). The dots stand for all interaction terms containing three or four heavy fields.

Now, one might presume that at least at the leading order of approximation, the effective action can be obtained simply by ignoring the heavy fields altogether and is given by the first line of the action~(\ref{action_exp}), which contains the kinetic terms for~$v_1$ and $\tilde{Y}$ and two interaction terms involving only these two light fields. As it happens, an action thus obtained would fail to capture properly the interactions of our low energy effective four-dimensional theory which, as we will see shortly,  are strongly modified by the presence of the heavy modes.  

The reason as for why this would have been the case, lies in the presence of trilinear interaction terms of the form $llh$ between the light modes ($l$) and the heavy modes ($h$), e.g.~$v_1^2u_n$  or $\partial^\mu\tilde{Y}\partial_\mu\tilde{Y}u_n$ appearing in the action~(\ref{action_exp}) through~$J_n^{(1)}u_n$.

\FIGURE{%
 \begin{pspicture}(-0.2,-1.2)(5.5,1.3)
\rput(0.0,-1){$l$}
\rput(0.0,1){$l$}
\rput(5.5,-1){$l$}
\rput(5.5,1){$l$}
\psline[linewidth=1pt,linestyle=solid](0.5,1)(2,0) 
\psline[linewidth=1pt,linestyle=solid](0.5,-1)(2,0)
\psline[linewidth=1pt,linestyle=solid](5,1)(3.5,0)
\psline[linewidth=1pt,linestyle=solid](5,-1)(3.5,0)
\psline[linewidth=3pt,linestyle=solid](1.95,0)(3.55,0)
\rput(2.7,-0.5){$h$} 
\end{pspicture}
\caption{Heavy mode correction to the quartic coupling of the light mode}
\label{llh} 
}  

It was already pointed out in~\cite{alberto}, in the context of 4D low energy effective actions derived from higher-dimensional theories with broken symmetries, that when terms of the type~$llh$ are present in the theory, the heavy modes cannot be simply dropped, as they will contribute to the effective action at the fourth order in fields through diagrams of the type depicted schematically on Fig.~\ref{llh}. What's more, due to the specific hierarchy of couplings between the light modes among themselves and their couplings to the infinite number of heavy fields (which is a characteristic feature of a theory obtained through a dimensional reduction) the contribution of the heavy fields towards the effective action is enhanced and is \emph{of the same order} as the interaction terms calculated using only the light modes. As a consequence, as we will see below, these interactions get quite drastically modified by the contributions coming from the heavy modes.

\subsection{Integrating out the heavy modes}
The procedure to derive the effective action for~$v_1$ and~$\tilde{Y}$ from~(\ref{action_exp}) is somewhat complicated by the fact that these two fields have very different couplings to the heavy modes, as well as  by the presence of the infinite towers of the heavy modes. It remains, however, quite standard in essence:
In order to obtain the correct effective action in the tree approximation, the heavy fields~$u_n$ and~$v_n$ must be \emph{integrated out}, which consists in solving their classical equations of motion:
\begin{eqnarray}
\partial^\mu\partial_\mu u_n +m_n^2 u_n&=&J_n^{(1)}+\sintfm (J^{(2)}_{nm}- \partial_\mu K^\mu_{mn})u_m\notag+ \\&&+\sintfm(K^{\mu}_{nm}-K^\mu_{mn})\partial_\mu u_m 
+\sinthm\tilde{\tilde{J}}_{nm}^{(2)}v_m+\dots
\label{eom_un} 
\end{eqnarray}\smallskip
\begin{eqnarray}
\partial^\mu\partial_\mu v_n +\tilde{m}_n^2 v_n&=&\tilde{J}_n^{(1)}+\sintfm (\tilde{J}^{(2)}_{nm}- \partial_\mu \tilde{K}^\mu_{mn})v_m+\notag\\&&+\sintfm(\tilde{K}^\mu_{nm}-\tilde{K}^\mu_{mn})\partial_\mu v_m + \sintfm\tilde{\tilde{J}}_{mn}^{(2)}u_m+\dots
\label{eom_vn}
\end{eqnarray}\medskip
 and then substituting the solutions back into the classical action~(\ref{action_exp}). 

Formally, the solutions of equations~(\ref{eom_un}) and~(\ref{eom_vn}), which we will denote~$\bar{u}_n$ and~$\bar{v}_n$, can be written:
\begin{eqnarray}
\bar{u}_n&=&\left(\partial^\mu\partial_\mu +m_n^2-J_{nn}^{(2)}+\partial_\mu K^\mu_{nn}\right)^{-1}\left[\vphantom{\sinthm}J_n^{(1)}+\smash{\sintfmneqn\left(J^{(2)}_{nm}- \partial_\mu K^\mu_{mn}\right)\bar{u}_m +}\right.\nonumber \\[2ex]
&&+\sintfmneqn\left(\vphantom{\tilde{K}^\mu_{mn}}K^{\mu}_{nm}-K^\mu_{mn}\right)\partial_\mu \bar{u}_m+\sinthm\tilde{\tilde{J}}_{nm}^{(2)}\bar{v}_m+\dots\left.\vphantom{\sintfm}\right]
\label{fsol_un} \end{eqnarray}
\begin{eqnarray}
\bar{v}_n&=&\left(\partial^\mu\partial_\mu +\tilde{m}_n^2-\tilde{J}_{nn}^{(2)}+\partial_\mu \tilde{K}^\mu_{nn}\right)^{-1}\left[\vphantom{\sinthm}\tilde{J}_n^{(1)}+\smash{\sinthmneqn \left(\tilde{J}^{(2)}_{nm}- \partial_\mu \tilde{K}^\mu_{mn}\right)\bar{v}_m+}\right.\nonumber \\[2ex]
&&+ \sinthmneqn\left(\tilde{K}^\mu_{nm}-\tilde{K}^\mu_{mn}\right)\partial_\mu \bar{v}_m +\sintfm\tilde{\tilde{J}}_{mm}^{(2)}\bar{u}_m+\dots\left.\vphantom{\sintfm}\right]\, .
\label{fsol_vn}
\end{eqnarray}
The explicit solutions are to be found perturbatively, in powers of~$1/m_n$ and~$1/\tilde{m}_n$, and of the couplings~$\lambda$,~$\alpha$ and~$\tilde{\lambda}$. 
Performing the expansion in~$1/m_n$ and~$1/\tilde{m}_n$ we obtain:
\begin{eqnarray}
\bar{u}_n&=&\left[\frac{1}{m_n^2}-\frac{1}{m_n^4}\partial^\mu\partial_\mu+\dots \right]J_n^{(1)}+\frac{1}{m_n^2}\sintfm\frac{1}{m_m^2}\left[J_{nm}^{(2)}-\partial_\mu K^\mu_{mn}\right]J_m^{(1)}+\nonumber\\
&&{}+\frac{1}{m_n^2}\sintfm\frac{1}{m_m^2}
\left(K^\mu_{nm}-K^\mu_{mn}\right)\partial_\mu J_m^{(1)} 
+\frac{1}{m_n^2}\sinthm \frac{1}{\tilde{m}_m^2}\tilde{\tilde{J}}^{(2)}_{nm}\tilde{J}_m^{(1)}+\dots\\
%\end{eqnarray}
%\begin{eqnarray}
 \bar{v}_n&=&\left[\frac{1}{\tilde{m}_n^2}-\frac{1}{\tilde{m}_n^4}\partial^\mu\partial_\mu+\dots \right]\tilde{J}_n^{(1)}+\frac{1}{\tilde{m}_n^2}\sinthm\frac{1}{\tilde{m}_m^2}\left[\tilde{J}_{nm}^{(2)}-\partial_\mu \tilde{K}^\mu_{mn}\right]\tilde{J}_m^{(1)}+\nonumber\\
&&{}+\frac{1}{\tilde{m}_n^2}\sinthm\frac{1}{\tilde{m}_m^2}
\left(\tilde{K}^\mu_{nm}-\tilde{K}^\mu_{mn}\right)\partial_\mu \tilde{J}_m^{(1)} 
+\frac{1}{\tilde{m}_n^2}\sintfm \frac{1}{{m}_m^2}\tilde{\tilde{J}}^{(2)}_{mn}{J}_m^{(1)}+\dots \ 
\end{eqnarray}   
Substituting the solutions~$\bar{u}_n$ and~$\bar{v}_n$ back into the action~(\ref{action_exp}) yields:
\begin{eqnarray}
S_{\text{eff}}&=&\int d^4 x \left\{-\tau+\frac12 \partial^\mu\tilde{Y}\partial_\mu\tilde{Y}+\frac12\partial^\mu v_1\partial_\mu v_1 -\frac12 \tilde{m}_1^2 v_1^2 \smash{ -\frac{{\lambda_4^{(0)}}\!\!\!}{4}\,v_1^4+\lambda_{(2,2)}^{(0)} v_1^2\,\partial^\mu\tilde{Y}\partial_\mu\tilde{Y} }\right. \nonumber \\ 
&&\hphantom{\int d^4 x' \left\{-\tau\right.}-\frac12\sintf J_n^{(1)}\left(\frac{1}{m_n^2}-\frac{1}{m_n^4}\partial^\mu\partial_\mu+\dots\right)J_n^{(1)} \nonumber\\
&&\hphantom{\int d^4 x' \left\{-\tau\right.}-\frac12\dsintf\frac{1}{m_n^2m_m^2}\left[\left(J_{nm}^{(2)}-\partial_\mu K^\mu_{mn}\right)J_m^{(1)}+\left(K^\mu_{nm}-K^\mu_{mn}\right)\partial_\mu J_m^{(1)}\right]J_n^{(1)} \nonumber \\
&&\hphantom{\int d^4 x' \left\{-\tau\right.}-\frac12\sinth \tilde{J}_n^{(1)}\left(\frac{1}{\tilde{m}_n^2}-\frac{1}{\tilde{m}_n^4}\partial^\mu\partial_\mu+\dots\right)\tilde{J}_n^{(1)} \nonumber\\
&&\hphantom{\int d^4 x' \left\{-\tau\right.}-\frac12\dsinth\frac{1}{\tilde{m}_n^2\tilde{m}_m^2}\left[\left(\tilde{J}_{nm}^{(2)}-\partial_\mu \tilde{K}^\mu_{mn}\right)\tilde{J}_m^{(1)}+\left(\tilde{K}^\mu_{nm}-\tilde{K}^\mu_{mn}\right)\partial_\mu \tilde{J}_m^{(1)}\right]\tilde{J}_n^{(1)} \nonumber \\
&&\hphantom{\int d^4 x' \left\{-\tau\right.}-\sintf\frac{1}{m_n^2}\,\sinthm\frac{1}{\tilde{m}_m^2}\tilde{\tilde{J}}_{mn}^{(2)}J_n^{(1)}\tilde{J}_m^{(1)} 
\left.\vphantom{\frac12}+\dots \right\} 
\end{eqnarray}
In order to determine the explicit form of the effective action in terms of~$v_1$ and~$\tilde{Y}$, we must use the explicit expressions for~$J_n^{(1)}$, $\tilde{J}_n^{(1)}$, $J_{nm}^{(2)}$, $\tilde{J}_{nm}^{(2)}$, $\tilde{\tilde{J}}_{nm}^{(2)}$, $K^\mu_{nm}$ and~$\tilde{K}^\mu_{mn}$. 
The dominant contributions to the action must come from~$(J^{(1)}/m_n)^2$ and~$(\tilde{J}_n^{(1)}/\tilde{m}_n)^2$, where~$J_n^{(1)}$ and~$\tilde{J}_n^{(1)}$ are given by the following combinations of the light fields:
\begin{eqnarray}
J_n^{(1)}&=&\frac{1}{\tau}{\left(\int dy\, \Phi_K'f_n'\right)}\partial^\mu\tilde{Y}\partial_\mu\tilde{Y}-\alpha{\left(\int dy\,\Phi_K h_1^2 f_n\right)} v_1^2 \label{current_1} \\
\tilde{J}_n^{(1)}&=&\frac{1}{\sqrt{\tau}}{\left(\int dy\, h_1h_n'\right)}{\left[-2\partial^\mu v_1\partial_\mu \tilde{Y}+v_1\partial^\mu\partial_\mu\tilde{Y}\right]} -\tilde{\lambda}{\left(\infint h_1^3 h_n\right)} v_1^3+\nonumber \\
&&{}+\frac{1}{\tau}{\left(\infint h_1'h_n'\right)}v_1\partial^\mu\tilde{Y}\partial_\mu\tilde{Y} \label{current_2}\ .
\end{eqnarray}
 Although the couplings involved in these expressions are fairly complicated functions of the background field and the wavefunctions of both light and heavy modes, we  can at least estimate their order of magnitude. The estimation can be performed very easily, through  dimensional analysis: each wavefunction contributes a factor~$a^{-1}$, each derivative gives a factor~$a$,~$\Phi_K \sim{\cal{O}}\left(a/\sqrt{\lambda}\right)$ and the tension is $\tau\sim{\cal{O}}\left(\lambda^{-1} a^3\right)$. Applying this simple recipe to~(\ref{current_1}) and~(\ref{current_2}) we see that:
\begin{eqnarray}
J_n^{(1)}&\sim &{\cal{O}}\left(\sqrt{\frac{\lambda}{a}}\right)\partial^\mu\tilde{Y}\partial_\mu \tilde{Y} +{\cal{O}}\left(\sqrt{\lambda a^3}\right)v_1^2 \label{current_est_1} \\
\tilde{J}_n^{(1)}&\sim & {\cal{O}}\left(\sqrt{\frac{\lambda}{a}}\right)\left[-2\partial^\mu v_1\partial_\mu \tilde{Y}+v_1\partial^\mu\partial_\mu\tilde{Y}\right] + {\cal{O}}\left(\tilde{\lambda} a\right) v_1^3+{\cal{O}}\left(\frac{\lambda}{a}\right)v_1\partial^\mu\tilde{Y}\partial_\mu\tilde{Y} \label{current_est_2}
\end{eqnarray} 
The contributions to the effective action relevant at the first order in~$\lambda$ can only come from the terms of~$J_n^{(1)}$ and~$\tilde{J}_n^{(1)}$ which are quadratic in~$v_1$ and~$\tilde{Y}$. It is also obvious that that the contributions coming from any terms involving~$J_{nm}^{(2)}$, $\tilde{J}_{nm}^{(2)}$, $\tilde{\tilde{J}}_{nm}^{(2)}$, $K^\mu_{nm}$ and~$\tilde{K}^\mu_{mn}$ are only relevant at order~$\lambda^2$, as they are all accompanied by factors~$\left(J^{(1)}/m_n\right)^2$ and~$(\tilde{J}_n^{(1)}/\tilde{m}_n)^2$ and therefore can be neglected.

\subsection{Effective action at the leading order}
Let us start by determining the effective action at order~$\lambda a^{-1}$, that is the order of the~$\lambda_{(2,2)}^{(0)}$ coupling between the branon and the scalar appearing in the action~(\ref{action_exp}).  
 
Eqns.~(\ref{current_est_1}) and~(\ref{current_est_2}) make apparent the fact which we have mentioned above, when discussing the importance of the heavy modes: the leading contributions to the quartic coupling~$\lambda_4 v_1^4$ and to the coupling~$\lambda_{(2,2)} v_1^2\partial^\mu\tilde{Y}\partial_\mu\tilde{Y} $ yielded by the heavy modes (that is, contributions involving the leading, constant term of the propagators) are of~${\cal{O}}\left(\lambda a\right)$ and of~${\cal{O}}\left(\lambda a^{-1}\right)$, respectively; that is~\emph{of the same order} as~$\lambda_4^{(0)}$ and~$\lambda_{(2,2) }^{(0)}$, given by~(\ref{lambda_4_0}) and~(\ref{lambda_22_0}). The actual values of the effective coupling constants can be found by performing the sums over the heavy modes (this can be done using completeness of the eigenfunctions and some algebraic relations between the background and the wavefunctions of the light modes; see Appendix~\ref{app_eff_couplings} for details). 

Once the couplings are calculated, the effective action at order~$\lambda a^{-1}$ reads:
 \begin{eqnarray}
S_{\text{eff}}&=&\int d^4 x \left\{-\tau+\frac12 \partial^\mu\tilde{Y}\partial_\mu\tilde{Y}+\frac12\partial^\mu v_1\partial_\mu v_1 -\frac12 \tilde{m}_1^2 v_1^2 \smash{-\frac{\lambda_4^{(1)}\!\!}{4}v_1^4
+ \frac{\lambda_4^\prime}{4} v_1^2\partial^\mu\partial_\mu v_1^2} \right\}, \label{action_leading_1}
\end{eqnarray}
where the contribution of the heavy modes shifted the effective quartic coupling constant:
 \begin{equation}
\lambda_4^{(1)}= \lambda_4^{(0)}
-{\frac{1}{2\tau^2}\sintf\frac{1}{m_n^2}\left(\infint \Phi_c h_1^2 f_n\right)^2}=\left(\tilde\lambda-\lambda\frac{\sigma^2}{4}\right)\infint h_1^4
\label{lambda_4_cor}
\end{equation}
and cancelled -- at the leading order $\lambda a^{-1}$ -- the coupling of the interaction term~$v_1^2\partial^\mu\tilde{Y}\partial_\mu\tilde{Y}$:  
\begin{equation}
\lambda^{(0)}_{(2,2)}-{\frac{\alpha}{\tau}\sintf\!\frac{1}{m_n^2}\!\left(\infint \Phi_K'f_n'\right)\!\!\left(\infint \Phi_Kh_1^2f_n\right)}=0\ .
\label{coupling_zero}
\end{equation}
The coupling~$\lambda_4^\prime$ is:
\begin{equation}
\lambda_4^\prime=\frac{\alpha^2}{2}\sintf\frac{1}{m_n^4}\left(\infint\Phi_K h_1^2 f_n\right)^2=\frac\lambda a J(\sigma)
\end{equation} 
with
$$
J(\sigma)=\frac{\sigma^2}{2}\left(\int_{-\infty}^{\infty}\frac{dy}{\cosh^{2\sigma}(y)}\right)^{-2} \int_{-\infty}^\infty \frac{dy}{\cosh^4(y)}\left(\int_0^ydy'\,[\cosh(ay')]^{2-2\sigma}\right)^2 \ .
$$
The action can be further simplified using 
$$
\int d^4x \, v_1^2\partial^\mu\partial^\mu v_1^2=-\frac43 \int d^4x \, v_1^3 \partial^\mu\partial_\mu v_1 = \frac43\tilde{m}_1^2\int d^4x \, v_1^4 \ ,
$$
where in the last step we have used the equation of motion for~$v_1$ (the quartic term is negligible at the first order in~$\lambda$). As the coupling $\lambda_4'\sim {\cal{O}}(\lambda a^{-1})$, this term therefore gives a correction to the quartic coupling suppressed by~$\tilde{m}_1^2/a^2$ and can be reabsorbed through the redefinition of this coupling: 
$$
\lambda_4=\lambda_4^{(1)}-\frac43\tilde{m}_1^2\lambda_4^\prime\approx \lambda_4^{(1)}\ .
$$ 
At the order~${\cal O}(\lambda a^{-1})$, our effective action reads therefore:
 \begin{equation}
S_{\text{eff}}=\int d^4 x \left\{-\tau+\frac12 \partial^\mu\tilde{Y}\partial_\mu\tilde{Y}+\frac12\partial^\mu v_1\partial_\mu v_1 -\frac12 \tilde{m}_1^2 v_1^2 -\frac{\lambda_4}{4}v_1^4 \right\}. 
\label{action_leading_2}
\end{equation}
Let us stress that the equations of motion for~$v_1$ and~$\tilde{Y}$ can be used (at order $\lambda a^{-1}$ or any higher order of the calculation) without any loss of generality and that, in particular, using them does not imply that our effective action is only valid on-shell. Any terms proportional to the equations of motion appearing at any given order of approximation in the calculation of the effective action, can be eliminated simply by a redefinition of the two fields which does not modify the rest of the action (for instance, the redefinition allowing to pass from~(\ref{action_leading_1}) to~(\ref{action_leading_2}) is~$v_1\to v_1 - \frac23 \lambda_4' v^3 $). Following this logic, in the subsequent calculations we can therefore also use the equation of motion for~$\tilde{Y}$ to set to zero all the terms of the effective action proportional to~$\partial^\mu\partial_\mu\tilde{Y}$.  

As we see, at order~${\cal{O}}(\lambda a^{-1})$, the branon is decoupled from~$v_1$. This  feature of the effective action~(\ref{action_leading_2}) is obviously in agreement with the Nambu-Goto action, eqn.~(\ref{NG_1}), in which any coupling of branon excitation to the matter  must be suppressed by the tension of the brane (which, let us remind,  is  in our case of order~${\cal{O}}(\lambda^{-1} a^{3})$).

\subsection{Effective action at order~${\cal{O}}(1/\tau)$}

Interactions between the scalar field~$v_1$ and the branon~$\tilde{Y}$ do appear at order~${\cal{O}}(\lambda a^{-3})$, or in other words at~${\cal{O}}\left(1/\tau\right)$. The effective action at this order reads~(the details of the calculation of coupling constants can be found in Appendix~\ref{app_eff_couplings}): 
\begin{eqnarray}
S_{\text{eff}}&=&\int d^4 x \left\{-\tau+\frac12 \partial^\mu\tilde{Y}\partial_\mu\tilde{Y}+\frac{1}{8\tau} \left(\partial^\mu\tilde{Y}\partial_\mu\tilde{Y}\right)^2+\frac12\partial^\mu v_1\partial_\mu v_1 -\frac12 \tilde{m}_1^2 v_1^2 -\frac{\lambda_4}{4}v_1^4 \right. \nonumber \\ 
 && \hphantom{\int d^4 x \left\{\right.}+\left.\frac{1}{2\tau}\partial^\mu\tilde{Y}\partial^\nu\tilde{Y}\partial_\mu v_1\partial_\nu v_1 +\frac{1}{2\tau} I(\sigma) \partial^\mu\tilde{Y}\partial_\mu\tilde{Y}\left[\partial^\nu v_1\partial_\nu v_1-\tilde{m}_1^2v_1^2\right]\right\}, \label{eff_action_simpl} 
\end{eqnarray} 
where $I(\sigma)$ is the following function:
\begin{equation}
\label{I(sigma)}
I(\sigma)\equiv\sigma \left({\displaystyle\int_{-\infty}^{\infty} dy \frac{1}{\cosh^{2\sigma}(x)} }\right)^{-1}{\displaystyle\int_{-\infty}^\infty dy \frac{y}{\cosh^4(y)}\int_0^y dy'\, [\cosh(y')]^{2-2\sigma}} \ .
\end{equation}

\FIGURE{%
\includegraphics[width=120mm,height=80mm]{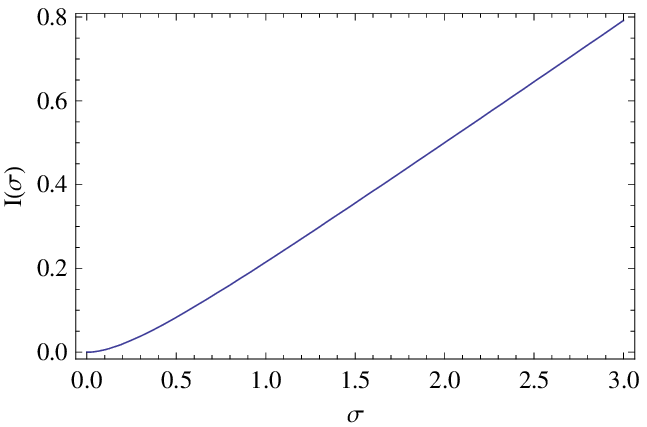}
\caption{Parameter $I(\sigma)$ appearing in the effective action}
\label{fIsigma}
}

The graph of the function~$I(\sigma)$ is presented on Fig.~\ref{fIsigma}.
Comparing our effective action, eqn.~(\ref{eff_action_simpl}), with the Nambu-Goto action, eqn.~(\ref{NG_exp}), we see that although the two are very similar, they are not identical. Indeed: 
\begin{equation}
S^{\text{scalar}}_{\text{NG}}-S_{\text{eff}}=  -\frac{1}{4\tau}\left(1+2I(\sigma)\right)\int d^4 x\, \partial^\mu\tilde{Y}\partial_\mu\tilde{Y}\left[\partial^\alpha v_1\partial_\alpha v_1-m_1^2v_1^2\right]\ .
\label{ng-eff}
\end{equation}
We see, therefore, that the actual form of the action seems to depend on the underlying field-theoretical model, even in the thin wall limit. Moreover, $I(\sigma)$ takes positive values for positive~$\sigma$, which means that whatever value of~$\sigma$ we choose, our effective action will \emph{never} be exactly equal to the Nambu-Goto action. 

\subsection{Covariant formulation}
Even if the effective action of the domain wall which we have determined is not identical to the Nambu-Goto action, we can expect it to obey the same fundamental covariance  principles, the branon~$\tilde{Y}$ entering the action only through the induced metric~$g_{\mu\nu}=\eta_{\mu\nu}-\partial_\mu\tilde{Y}\partial_\nu\tilde{Y}/\tau$ (and its derivatives). We should therefore be able to rewrite our effective action~(\ref{eff_action_simpl}) in terms of geometric invariants, as it is the case for the Nambu-Goto action. This amounts to rewriting~(\ref{ng-eff}) in a covariant way, which can be achieved noticing that (up to total derivatives and inessential terms involving derivatives of~$\partial^\alpha\partial_\alpha\tilde{Y}$):
\begin{eqnarray*}
S^{\text{scalar}}_{\text{NG}}-S_{\text{eff}}&=&  -\frac{1}{8\tau}\left(1+2I(\sigma)\right)\int d^4 x\, \partial^\mu\tilde{Y}\partial_\mu\tilde{Y}\partial^\alpha \partial_\alpha v_1^2 \nonumber \\
&=& -\frac{1}{4\tau}\left(1+2I(\sigma)\right)\int d^4 x\, v_1^2 \, \partial^\alpha\partial^\mu\tilde{Y}\partial_\alpha\partial_\mu\tilde{Y} \ .
\end{eqnarray*}
This last expression can be readily related with the Ricci scalar calculated from the induced metric, which reads:
\begin{equation}
R=-\frac{1}{\tau}\partial^\alpha\partial_\alpha{\left(\partial^\nu\tilde{Y}\partial_\nu\tilde{Y}\right)}+\frac{1}{\tau}\partial^\alpha\partial_\nu{\left(\partial_\alpha\tilde{Y}\partial^\nu\tilde{Y}\right)}=-\frac{1}{\tau}\partial^\alpha\partial^\nu\tilde{Y}\partial_\alpha\partial_\nu\tilde{Y}+\frac{1}{\tau}{\left(\partial^\nu\partial_\nu\tilde{Y}\right)}^2 \ .
\label{R}
\end{equation}
Our effective action can therefore be written:
\begin{equation}
S_{\text{eff}}=\int d^4 x \sqrt{g}\left\{-\tau
+\frac12 g^{\mu\nu}\partial_\mu v_1\partial_\nu v_1 -\frac12 \tilde{m}_1^2 v_1^2 -\frac{\lambda_4}{4}v_1^4 -\frac14\left(1+ 2I(\sigma)\right) v_1^2 R \right\}\label{eff_action_geom}\, ,
\end{equation}
where we have eliminated the inessential term~$(\partial^\nu\partial_\nu\tilde{Y})^2$ appearing in the expression for~$R$. The action~(\ref{eff_action_geom}) is  therefore the sum of the familiar Nambu-Goto action and an additional interaction term between the scalar and the branon involving the Ricci scalar of the induced metric~$\xi v_1^2\, R$. Our model of domain wall being fairly generic, the presence of this term is also a generic feature of the brane action. 

Non-minimal coupling of the form~$\xi R v_1^2$ reminds the gravitational term in tensor-scalar theories of gravity~\cite{bd}, appearing naturally in string theory as dilaton coupling. It should be stressed, however, that there is no dynamical graviton in our setup and that the curvature~$R$ is here only a self-coupling of the branon~$\tilde{Y}$. It is quite likely though  that  such terms, with~$R$ playing the role of the kinetic term of the graviton, would be generated as induced gravity effects~\cite{dgp_1} when quantum corrections are taken into account.        

\subsection{Finite-width corrections}
To conclude this section, let us take a brief look at the corrections of order~${\cal{O}}\left(\lambda a^{-5} \right)$ to our effective action, focusing on the branon part. At this order the brane action reads:
\begin{eqnarray}
S_{\text{brane}}&\approx&\int d^4 x \left\{-\tau+\frac12\partial^\mu\tilde{Y}\partial_\mu\tilde{Y}+ \right.\nonumber \\
&&\quad{}+ \frac{1}{2\tau^2}\sintf \left(\infint \Phi_K' f_n'\right)^2 \partial^\mu\tilde{Y}\partial_\mu\tilde{Y}\left[\frac{1}{m_n^2}-\frac{1}{m_n^4}\partial^\alpha\partial_\alpha\right]\partial^\nu\tilde{Y}\partial_\nu\tilde{Y}\left.\vphantom{\frac12}\right\}.
\end{eqnarray}
Calculating the contribution of ${\cal O}(p^2)$ corrections in the expansion of the propagators we find:       \begin{equation}
S_{\text{brane}}=\int d^4 x {\left\{-\tau +\frac12\partial^\mu\tilde{Y}\partial_\mu\tilde{Y}+\frac{1}{8\tau} {\left(\partial^\mu\tilde{Y}\partial_\mu\tilde{Y}\right)}^2-\frac{\pi^2-6}{48\tau a^2}{\left(\partial^\mu\tilde{Y}\partial_\mu\tilde{Y}\right)}{\left(\partial^\alpha\partial^\nu\tilde{Y}\partial_\alpha\partial_\nu\tilde{Y}\right)}\right\}}.
\label{action_brane_width}
\end{equation} 
In the last term of this action, a subdominant quartic self-coupling of~$\tilde{Y}$, we can again identify the curvature scalar~$R$. As~$R$ is a total derivative (see eqn.~(\ref{R})),  
 we can include it in the action~(\ref{action_brane_width}), which can then be rewritten in the covariant form:
\begin{equation}
S_{\text{brane}}=\int d^4 x\,\sqrt{g} \left\{-\tau -\frac{\pi^2-6}{24 a^2}R\tau \right\}\, .
\end{equation}
This is the correction to the Nambu-Goto action which has been obtained in~\cite{ruth_carter} using Gauss-Coddazzi formalism in the case when the width of the domain wall (which, let us remind, is of order ${\cal{O}}(a^{-1})$) is small but finite. The ``antirigidity'' term proportional to the trace~$K$ of the extrinsic curvature appearing in~\cite{ruth_carter} (see eqn.~(\ref{width_corrs})) is absent in our setup. Also, any terms proportional to~$K$ appearing at  higher orders in the calculation of the effective action can be set to zero. Indeed, in terms of the branon field, the extrinsic curvature reads~\mbox{$K_{\mu\nu}=\partial_\mu\partial_\nu \tilde{Y}/\sqrt{\tau}$} and therefore~$K=K^\mu_\mu$ is proportional to the lowest order equation of motion for~$\tilde{Y}$, \mbox{$K=\partial^\mu\partial_\mu \tilde{Y}/\sqrt{\tau}$}. As we have already stressed, within the region of validity of our effective theory, any such terms can be consistently eliminated from the action through redefinition of the fields. 

\section{Physical implications\label{section_physical_implications}}

We have seen that our low energy effective action differs from the Nambu-Goto action by an additional interaction term of the scalar field~$v_1$ with the branon~$\tilde{Y}$. Naturally, we expect the presence of this term to have physical effects and in this section we will investigate some of them. 

\subsection{Scalar-branon scattering}
The simplest physical process in which we expect a manifestation of the difference between our effective action and Nambu-Goto action is the scattering of the scalar field~$v_1$ on branons. At leading order, it is described by the following terms in the action (\ref{eff_action_simpl}):
\beq
\frac{1}{2\tau}\partial_\mu \tilde{Y} \partial^\nu \tilde{Y} \partial_\nu v_1\partial^\mu v_1+
\frac{1}{2\tau}I(\sigma)\partial_\mu \tilde{Y} \partial^\mu \tilde{Y} \partial^\nu v_1\partial_\nu v_1-
\frac{1}{2\tau}I(\sigma)\tilde{m}_1^2v_1^2\partial_\mu \tilde{Y} \partial^\mu \tilde{Y} 
\eeq
These give a four-particle vertex with amplitude
\bea
\label{vertex}
g(p_1,p_2,p_3,p_4)= 
\begin{pspicture}(0,0)(4,1.2)
\rput(0.2,-1){$v_1$}
\rput(0.2,1){$v_1$}
\rput(1.2,1){$\vec p_1$}
\rput(1.2,-0.2){$\vec p_2$}
\rput(3.8,-1){$\tilde{Y}$}
\rput(3,-0.3){$\vec p_3$}
\rput(3,1){$\vec p_4$}
\rput(3.8,1){$\tilde{Y}$}
\psline[linewidth=1.5pt,linestyle=solid](0.5,1)(2,0) 
\psline[linewidth=1.5pt,linestyle=solid](0.5,-1)(2,0) 
\psline[linewidth=1.5pt,linestyle=dashed](2,0)(3.5,-1)
\psline[linewidth=1.5pt,linestyle=dashed](2,0)(3.5,1)
\end{pspicture}
&=&-i\frac{1}{\tau}\left[p_{2,\mu}p_4^\mu p_{1,\nu}p_3^\nu+p_{1,\mu}p_4^\mu p_{2,\nu}p_3^\nu \right]\notag\\&&
-i\frac{2}{\tau}I(\sigma)p_{3,\nu}p_4^\nu\left[p_{1,\mu}p_2^\mu+\tilde{m}_1^2\right]
,\eea
where we consider that all momenta are incoming. In the center-of-mass frame we have, 
 $p_1=(\omega_1,\vec p_1), p_2=(\omega_1,-\vec p_1)$ with $\omega_1=\sqrt{|\vec p_1|^2+\tilde{m}_1^2}$ and $p_3=(\omega_3,\vec p_3), p_4=(\omega_3,-\vec p_3)$ with $\omega_3=|\vec p_3|$.

The differential cross section for this process is:
\beq
\frac{d\sigma}{d\Omega}=\frac{1}{2(8\pi)^2 s}\frac{|\vec p_3|}{|\vec p_1|}|g|^2\ ,
\eeq
where 
$$
|g|^2=\frac{1}{\tau^2}\left|2\omega_1^2\omega_2^2(1+4I(\sigma))+2(\vec p_1\cdot \vec p_3)^2\right|
=\frac{1}{(8\tau)^2}\left|(2\tilde{m}_1^2-t-u)2(1+4I)+(t-u)^2\right|^2\ ,
$$
$s=(p_1+p_2)^2=4w_1^2,\ t=(p_1-p_3)^2,\ u=(p_1-p_4)^2$ being the Mandelstam variables.  

The corresponding result for the Nambu-Goto action is obtained when replacing $I(\sigma)$ by~$-\frac 1 2$. Remember however that $I(\sigma)$ never reaches $-\frac 1 2$ for any $\sigma$. 

\subsection{Corrections to the potential of the fifth force}

As it has already been noticed in~\cite{kugo_ng}, massless branons can mediate a long-range force
between particles on the brane. In~\cite{kugo_ng} the static potential arising from the branon exchange between two fermions of masses $m$ and~$m'$ has been derived:
\begin{equation}
V(r)=-\frac{3}{128\pi^3\tau^2}\frac{mm'}{r^7}\ ,
\label{fermion_pot}
\end{equation}
where $r$ denotes the distance between the two particles and the minus sign implies an attractive force. Given that branons couple to matter fields only through the induced metric, we can expect this force to be gravity-like  in the sense to be universal (independent of charge or spin of the interacting particles, as suggested by~(\ref{fermion_pot})). We expect the additional coupling between the light scalar and the branon appearing in our effective action to produce corrections to the potential~(\ref{fermion_pot}). To determine the nature of these corrections, let us calculate  the potential resulting from the branon exchange between two scalars~$v_1$.  We follow here the analysis of~\cite{kugo_ng}, where more details on the calculation can be found.

The leading contribution to the static potential between two~$v_1$'s of mass~$\tilde{m}_1$ comes from the one loop diagram containing two vertices (\ref{vertex}):
\bea
-i{\cal{M}}_{\text{eff}}(p_1,p_3,q)&=&\begin{pspicture}(-0.5,0)(5,1.2)
\rput(0.2,-1){$v_1$}
\rput(0.2,1){$v_1$}
\rput(1.9,-0.2){$\tilde{Y}$}
\rput(3.8,-0.2){$\tilde{Y}$}
\rput(0.8,0.7){$\vec p_1$}
\rput(0.8,-0.7){$\vec p_2$}
\rput(4.8,-1){$v_1$}
\rput(4.3,0.7){$\vec p_3$}
\rput(4.3,-0.7){$\vec p_4$}
\rput(3.65,0.3){$\vec k$}
\rput(2.2,0.3){$\vec q-\vec k$}
\rput(4.8,1){$v_1$}
\psline[linewidth=1.5pt,linestyle=solid](0.5,1)(4.5,1) 
\psline[linewidth=1.5pt,linestyle=solid](0.5,-1)(4.5,-1) 
\pscircle[linewidth=1.5pt,linestyle=dashed](2.5,0){1} 
\end{pspicture}
\notag \\
\notag \\
\notag \\
&=&\frac{1}{2\tau^2}\int \frac{d^4k}{(2\pi)^4}\frac{g(p_1,p_2,-k,k-q)g(p_3,p_4,k,q-k)}{k^2(k-q)^2}
\eea
The scattering  amplitude~${\cal{M}}_{\text{eff}}$ can be computed using the method of dimensional regularization (by changing the dimension in the integral to $n=4-2\varepsilon$ to extract divergent contributions). When the scalars are non-relativistic ($p_1=p_3=(\tilde{m}_1,\vec{0})$), it depends on the momentum transfer $q$ only,
\bea
-i{\cal{M}}_{\text{eff}}(q)&=&\frac{1}{2\tau^2}\left[\tilde{m}_1^4q^4\left(\frac{23}{150}-\frac{1}{20}\log\left(-\frac{q^2}{\mu^2}\right)+\frac{1}{20\varepsilon}\right)\right.\notag\\
&&\quad-\left.\tilde{m}_1^2q^6 I(\sigma)\left(\frac{4}{9}-\frac{1}{6}\log\left(-\frac{q^2}{\mu^2}\right)+\frac{1}{6\varepsilon}\right)\right.\\
&&\quad+\left.q^8I^2(\sigma)\left(\frac{1}{2}-\frac{1}{4}\log\left(-\frac{q^2}{\mu^2}\right)+\frac{1}{4\varepsilon}\right)\right]\notag \ ,
\eea
where we have introduced a renormalization scale~$\mu$. The divergent constants can be renormalized introducing higher-dimensional counterterms in the Lagrangian, only the logarithmic terms are of physical relevance. The effective static potential can be computed as the Fourier transform of the amplitude~${\cal{M}}_{\text{eff}}$:
\beq
V(r)=\frac{1}{4\tilde{m}_1^2}\int \frac{d^3q}{2\pi^3}e^{iqr}{\cal{M}}_{\text{eff}}(q),
\eeq
where the factor $4\tilde{m}_1^2$ comes from the difference of normalization between relativistic and non-relativistic fields.
The integral can be performed in the complex plane and only the discontinuity of the logarithm contributes. The static potential reads:
\beq
V(r)=-\frac{3}{128 \pi^3 \tau^2}\frac{\tilde{m}_1^2}{r^7}-\frac{105 I(\sigma)}{8 \pi^3 \tau^2}\left[\frac{1}{4r^9}+\frac{27 I(\sigma)}{\tilde{m}_1^2 r^{11}}\right].
\eeq
The leading term is identical to~(\ref{fermion_pot}). As expected, non-minimal coupling of the scalar to the branon field produces subleading contributions to the static potential. It should be stressed that as the  coupling~$\xi R v_1^2$ is particular to scalars (for dimensional reasons), we cannot expect these 
corrections to be universal.

\subsection{``Higgs'' decay into branons}

The most remarkable effect of the~$\xi Rv_1^2$ coupling in our effective action appears in the presence of spontaneous symmetry breaking in the four-dimensional theory\footnote{Strictly speaking our derivation of the effective action is not valid in this case, as when~$\tilde{m}_1^2<0$ the kink configuration~(\ref{kink}) becomes unstable. However, by covariance principles, we can expect that the form of the action will be the same (except for the sign of the mass term). This is confirmed by our calculations performed  by perturbation around the background configuration in the broken phase, which will be presented elsewhere.}. Indeed, it can be readily seen that this term opens the possibility of decay of the scalar~$v_1$ (which we can now consider as a toy Higgs field) into branons.
When changing the sign of the mass term in our effective action, eqn.~(\ref{eff_action_geom}),~$v_1$ acquires a vacuum expectation value~$c_v\approx \tilde{m}_1^2/\lambda_4$. Shifting the field, $v_1\to v_1+c_v$, we obtain the following trilinear interaction term in the Lagrangian:
\beq
\frac{\tilde{m}_1^2}{2\tau}\left(I(\sigma)+\frac 1 2\right)v_1^2\partial_\mu \tilde{Y} \partial^\mu \tilde{Y}\xrightarrow{\,v_1\to v_1+c_v\,} \frac{c_v}{\tau}\tilde{m}_1^2\left(I(\sigma)+\frac 1 2\right)v_1\partial_\mu \tilde{Y} \partial^\mu \tilde{Y}+\dots \ ,
\eeq
which gives a three-particle vertex with amplitude
\beq
\mathcal{M}= 
\begin{pspicture}(0,0)(4,1)
\rput(0.5,-0.3){$v_1$}
\rput(1,0.3){$\vec q$}
\rput(4,-1){$\tilde{Y}$}
\rput(3,-0.3){$\vec k$}
\rput(3,1){$\vec p$}
\rput(4,1){$\tilde{Y}$}
\psline[linewidth=1.5pt,linestyle=solid](0.5,0)(2,0) 
\psline[linewidth=1.5pt,linestyle=dashed](2,0)(3.5,-1)
\psline[linewidth=1.5pt,linestyle=dashed](2,0)(3.5,1)
\end{pspicture}
=\frac{c_v}{\tau}\left(I(\sigma)+\frac 1 2\right)\tilde{m}_1^2k_\mu p^\mu.
\vspace{7mm}
\eeq
This yields the rate of the decay of~$v_1$ for into two branons: 
\beq
\Gamma_{v_1\to \tilde{Y}\tilde{Y}}=\frac{1}{4\tilde{m}_1}\int \frac{d^3k}{(2\pi)^3}\frac{d^3p}{(2\pi)^3}\frac{(2\pi)^4\delta^4(q-p-k)}{4p_0k_0}|\mathcal{M}|^2.
\eeq
If the Higgs particle is at rest, we have $q=(\tilde{m}_1,0), k=(\frac{\tilde{m}_1}{2},\vec k), p=(\frac{\tilde{m}_1}{2},-\vec k)$ with $|\vec k|=\frac{\tilde{m}_1}{2}$. In this case, the decay rate is 
\beq
\Gamma_{v_1\to \tilde{Y}\tilde{Y}}=\frac{\tilde{m}_1^7c_v^2}{128\pi\tau^2}\left(I(\sigma)+\frac 1 2\right)^2=\frac{\tilde{m}_1^9}{256\pi\lambda_4\tau^2}\left(I(\sigma)+\frac 1 2\right)^2.
\eeq
It is natural to assume that in a more realistic brane model, involving the physical Higgs, such a process would also be present. It could be a good indicator of the existence of branons and would put bounds on the allowed value of the brane tension. For instance, if the Higgs mass is 150 GeV and the brane tension is~$\tau=\,$(300GeV)$^4$, then the Higgs decay width into branons is~$\Gamma_{v_1\to \tilde{Y}\tilde{Y}}\approx 4\cdot 10^{-3}$ GeV which is roughly 1/3 of the total decay width for the Higgs in the Standard Model~(through standard, non-exotic, channels), see~\cite{branons_lhc}. Of course, for these optimistic values of~$\tilde{m}_1$ and~$\tau$ our calculations are only marginally valid and for smaller ratios of the Higgs mass to the tension, this decay is rapidly suppressed (because of the factor~$(\tilde{m}_1^4/\tau)^2$)~and therefore very difficult to detect.

\section{Conclusions\label{section_conclusions}}
We have determined the effective action on the brane in an explicit field-theoretical model, where the brane is modelled by a domain wall in the five-dimensional Minkowski space. In this model, the four-dimensional low energy field theory contains two scalar fields localized on the brane, one of which is a massless branon excitation whereas the other (endowed with a small mass) plays the role of a matter field. Our calculation confirms the importance of the heavy modes of the spectrum in the derivation of the effective action, as pointed out in~\cite{alberto}. The comparison of the four-dimensional effective action with the Nambu-Goto action has revealed the presence of an additional interaction term between the branon and the light scalar, which is proportional to the curvature scalar. Presence of this  term modifies the cross section of the branon-scalar scattering and yields short-distance corrections to  the potential of the ``fifth force'' mediated by branons. In the presence of spontaneous symmetry breaking, this term allows for a non-zero decay of the ``Higgs'' into branons, which could be used to put bounds on the allowed value of the brane tension.         

\acknowledgments It is a pleasure to thank M.~Shaposhnikov, R.~Gregory, K.~Tamvakis and A.~Salvio for valuable discussions. The work of KZ is supported by the EU FP6 Marie Curie Research and Training Network ``UniverseNet'' (MRTN-CT-2006-035863). KZ would also like to thank the Dept. of Mathematical Sciences of Durham University, where part of this work was performed, for its hospitality and the Swiss National Science Foundation for the financial support during this stay (grant PBEL2-114438). 

\appendix
\section{Explicit expressions for the couplings of the heavy fields\label{app_heavy_couplings}}
The explicit expressions for the combinations of lights fields coupling to the heavy fields, entering the action~(\ref{action_exp}) are:
\begin{eqnarray}
J_n^{(1)}&=&\frac{1}{\tau}\left(\int dy\, \Phi_K'f_n'\right)\,\partial^\mu\tilde{Y}\partial_\mu\tilde{Y}-\alpha\left(\int dy\,\Phi_K h_1^2 f_n\right)\, v_1^2  \\
\tilde{J}_n^{(1)}&=&\frac{1}{\sqrt{\tau}}\left(\int dy\, h_1h_n'\right) \left[-2\partial^\mu v_1\partial_\mu \tilde{Y}+v_1\partial^\mu\partial_\mu\tilde{Y}\right] -\tilde{\lambda}\left(\infint h_1^3 h_n\right) v_1^3+\nonumber\\
&&{}+\frac{1}{\tau}\left(\infint h_1'h_n'\right)v_1\partial^\mu\tilde{Y}\partial_\mu\tilde{Y}\\
J_{nm}^{(2)}&=&-\alpha\left(\infint h_1^2f_nf_m\right)v_1^2+\frac{1}{\tau}\left(\infint f_n'f_m'\right)\partial^\mu\tilde{Y}\partial_\mu\tilde{Y}\\ 
K^\mu_{nm}&=&-\frac{1}{\sqrt\tau}\left(\infint f_n'f_m\right)\partial^\mu\tilde{Y}   \\
\tilde{J}^{(2)}_{nm}&=&  -3\tilde{\lambda}\left(\infint h_1^2h_nh_m\right)v_1^2+\frac{1}{\tau}\left(\infint h_n'h_m'\right)\partial^\mu\tilde{Y}\partial_\mu\tilde{Y}\\ 
\tilde{K}^\mu_{nm}&=&-\frac{1}{\sqrt\tau}\left(\infint h_n'h_m\right)\partial^\mu\tilde{Y}   \\
\tilde{\tilde{J}}^{(2)}_{nm}&=&-\alpha\left(\infint \Phi_Kh_1f_nh_m\right) v_1
\label{currents}
\end{eqnarray}
\section{Calculation of the effective couplings\label{app_eff_couplings}}
The technique to derive the effective couplings of our effective action~(\ref{action_exp}) is conceptually quite simple (although in practice its implementation is not always entirely straightforward). The difficult part is, of course, to  calculate the sums over the heavy modes appearing in the expressions for the couplings. The basic idea is to cancel the~$1/m_n^2$ factor entering the expressions under the sum by extracting a factor~$m_n^2$ from one of the integrals under the sum  (when calculating higher order corrections, one needs to extract the necessary number of such factors). This can be achieved with the help of the following identity 
\begin{eqnarray}
\label{identity}
F'(y)\partial_r=\frac{1}{2}[F(y),\Omega^2]-\frac{1}{2} F''(y)\ .
\end{eqnarray}
Here $F(y)$ is an arbitrary function and $\Omega^2=-\partial_y^2+U(y)$ is a self-adjoint Schr\"odinger operator with some potential~$U(y)$. We will use the Schr\"odinger operators determining the spectrum of our theory, that is either~$\Omega^2_f f_n=m_n^2 f_n$ or~$\Omega^2_h h_n=\tilde{m}_n^2h_n$.
Eliminating the dependence on the masses allows us to use the completeness of the wavefunctions to perform the sum over the modes and end up with closed-form expressions for the effective couplings.
 
As a concrete application of this method, let us present the calculation of the quartic self-interaction for~$v_1$: 
\begin{equation}\label{lambda_4}
\lambda_4\equiv \frac{\tilde{\lambda}}{4}\infint h_1^4-{\frac{1}{2\tau^2}\sintf\frac{1}{m_n^2}\left(\infint \Phi_K h_1^2 f_n\right)^2}\ .
\end{equation}
We begin by deriving the equality
\begin{equation}\label{eq2}
\infint \Phi_K h_1^2 f_n=\frac{m_n^2v}{2a(1+\sigma)}\infint f_0 F f_n\ , \qquad \mbox{where}\qquad F'=\frac{h_1^2}{f_0}\ .
\end{equation} 
To start, we rewrite the integral as follows:
$$
\infint\Phi_Kh_1^2f_n =\infint\Phi_Kf_0\frac{h_1^2}{f_0}f_n=-\frac{v}{2a}\infint f_0'\frac{h_1^2}{f_0}f_n=-\frac{v}{2a}\infint f_nF'f_0'\ ,
$$
where we have set~$F'={h_1^2}/{f_0}$. Using the identity~(\ref{identity}) we obtain:
\begin{eqnarray*}
\infint\Phi_Kh_1^2f_n &=& -\frac{v}{2a}\infint f_nF'f_0'=-\frac{v}{4a}\infint f_n\left\{[F,\Omega_f^2]-F''\right\}f_0\\[.5ex]
&=&-\frac{v}{4a}\left\{-m_n^2\infint f_nFf_0 -\frac{2a}{v}(1-\sigma)\infint \Phi_Kf_nF'f_0\right\} \\[.5ex]
&=&m_n^2\frac{v}{4a}\infint f_nFf_0+\frac{1}{2}(1-\sigma)\infint\Phi_Kh_1^2f_n \ ,
\end{eqnarray*}
where we have used~$F''=(2a/v)(1-\sigma)\Phi_KF'$.
This yields the identity~(\ref{eq2}).

\noindent We can now perform the sum over modes in the expression for the effective coupling~$\lambda_4$, eqn.~(\ref{lambda_4}). Indeed, using~(\ref{eq2}) we have:

\begin{eqnarray*}
\frac{\alpha^2}{2}\sintf\frac{1}{{m}_n^2}{\left(\infint\Phi_K h_1^2f_n\right)}^2&=&\frac{\alpha^2}{2}\sintf\frac{1}{{m}_n^2}{\left(\infint\Phi_K h_1^2f_n\right)}{\left(\frac{v}{2a(\sigma+1)}m_n^2\infint f_0Ff_n\right) }\\ [.5ex]
&=&\frac{\alpha^2v}{4a(\sigma+1)}\infint\!\tinfint \Phi_K(y) h_1^2(y) f_0(\tilde{y}) F(\tilde{y})\sintf \!f_n(y)f_n(\tilde{y})
\end{eqnarray*}
Using the completeness relation:
$$\sintf f_n(y)f_n(\tilde{y})=\delta(y-\tilde{y})-f_0(y)f_0(\tilde{y})$$
 we obtain:
$$
\frac{\alpha^2}{2}\sintf\frac{1}{{m}_n^2}\left(\infint\Phi_K h_1^2f_n\right)^2=\frac{\alpha^2v}{4a(\sigma+1)}\infint\Phi_Kh_1^2 f_0 F \ .
$$
(Given that $\Phi_K$ is odd and $h_1$ even, the zero-mode term does not contribute ). Upon using $\alpha=\lambda\sigma(\sigma+1)/2$ and the identity~
\begin{equation}
\Phi_Kh_1^2f_0=-\frac{v}{2a(\sigma+1)}(h_1^2f_0)'
\end{equation}
this  can be further rewritten as:
$$
\frac{\alpha^2}{2}\sintf\frac{1}{{m}_n^2}\left(\infint\Phi_K h_1^2f_n\right)^2=-\frac{\alpha^2v^2}{8a(\sigma+1)^2}\infint(h_1^2 f_0)' F
=\lambda\frac{\sigma^2}{16}\infint h_1^4  \ .
$$
Inserting this result into eqn.~(\ref{lambda_4}) we obtain, as advertised:
$$
\lambda_4=\left(\frac{\tilde\lambda}{4}-\lambda\frac{\sigma^2}{16}\right)\infint h_1^4  \ .
$$
As all the other couplings present in our effective action can be calculated in a similar manner, we will skip the details of their calculation and will instead present the summary of the results together with the equalities of the type of~(\ref{eq2}) which we have used. 

\noindent The identity :
\begin{equation}
\infint \Phi_K'f_n'= \frac12 m_n^2\infint y\, \Phi_K'f_n \ , 
\label{eq3}
\end{equation} 
(which was used in~\cite{gervais} in the derivation of the perturbation theory for the one-dimensional kink) allows us to derive: 
\begin{equation}
\sintf\! \frac{1}{m_n^2}\left(\infint \Phi_K'f_n'\right)^2=-\frac12 \infint y\,\Phi'_K\Phi''_K =\frac14\tau
\end{equation}
\begin{equation}
\sintf\! \frac{1}{m_n^4}\left(\infint \Phi_K'f_n'\right)^2=\infint y^2\Phi_K'^2=\frac{\tau}{48a^2}\left(\pi^2-6\right)
\end{equation}

\begin{equation}
\sintf\!\frac{1}{m_n^2}\!\!\left(\infint \Phi_K'f_n'\right)\!\!\!\left(\infint \Phi_K h_1^2f_n\right)=\frac{\sigma v}{2(1+\sigma)(1+2\sigma)}
\end{equation}
Combining~(\ref{eq2}) with~(\ref{eq3}) yields:
\begin{equation}
\sintf\!\frac{1}{m_n^4}\!\!\left(\infint \Phi_K'f_n'\right)\!\!\!\left(\infint \Phi_K h_1^2f_n\right)=\frac{v^2}{4\sigma(1+\sigma)a^2}I(\sigma)
\end{equation}
where
\begin{equation}
I(\sigma)\equiv\sigma \left({\displaystyle\int_{-\infty}^{\infty} dy \frac{1}{\cosh^{2\sigma}(x)} }\right)^{-1}{\displaystyle\int_{-\infty}^\infty dy \frac{y}{\cosh^4(y)}\int_0^y dy'\, [\cosh(y')]^{2-2\sigma}} \ .
\end{equation}
\noindent Using twice the identity~(\ref{eq2}) we obtain:
\begin{equation}
\sintf\frac{1}{m_n^4}\left(\infint\Phi_K h_1^2 f_n\right)^2=\frac1a J(\sigma)
\end{equation} 
with
$$
J(\sigma)\equiv\frac{\sigma^2}{2}\left(\int_{-\infty}^{\infty}\frac{dy}{\cosh^{2\sigma}(y)}\right)^{-2} \int_{-\infty}^\infty \frac{dy}{\cosh^4(y)}\left(\int_0^ydy'\,[\cosh(ay')]^{2-2\sigma}\right)^2 \ .
$$
\noindent Finally, the identity: 
\begin{equation}
\infint h_1h_n'= \frac12\left(\tilde{m}_n^2-\tilde{m}_1^2\right)\infint y\, h_1 h_n
\end{equation}
allows us to derive:
\begin{equation}
\sinth\! \frac{1}{\tilde{m}_n^2}\left(\infint h_1h_n'\right)^2\approx\frac14-\frac{\tilde{m}_1^2}{2}\infint y^2 h_1^2\approx \frac14
\end{equation}

\bibliographystyle{JHEP}

\providecommand{\href}[2]{#2}\begingroup\raggedright\endgroup

\end{document}